\documentstyle[prl,aps,floats,psfig]{revtex}

\begin{document}
\twocolumn[
\hsize\textwidth\columnwidth\hsize\csname@twocolumnfalse\endcsname

\draft

\title{
Polaron and bipolaron formation in a cubic perovskite lattice
}
\author{Vladimir N. Kostur and Philip B. Allen}
\address{
Department of Physics, State University of New York,
Stony Brook, NY 11794-3800
}
\maketitle
%\receipt{   }
\date{\today}
%
% This paper has 7 figures.
%	Fig1: \label{fig:nocoul} radius,energy,gap vs coupling constant
%	Fig2: \label{fig:sizerr} finite size errors
%	Fig3: \label{fig:LP} energy versus radius of Landau-Pekar approx
%	Fig4: \label{fig:vib} vibrational energies versus coupling const
%	Fig5: \label{fig:2en} Bipolaron energy
%	Fig6: \label{fig:2rad} Bipolaron radius
%	Fig7: \label{fig:2gap} Bipolaron gap
%
% This paper has labeled equations:
%	\label{eq:Hamiltonian}
%	\label{eq:hopping}
%	\label{eq:elph}
%	\label{eq:bareph}
%	\label{eq:coulomb}
%	\label{eq:1elpsi}
%	\label{eq:selfen}
%	\label{eq:selfen1}
%	\label{eq:selfensum}
%	\label{eq:sinsum}
%	\label{eq:location}
%	\label{eq:radius}
%	\label{eq:LP}
%	\label{eq:enLP}
%	\label{eq:radLP}
%	\label{eq:chspr}
%	\label{eq:LPen}
%	\label{eq:LPspr}
%	\label{eq:2elpsi}
%	\label{eq:2rad}
%	
% %
\begin{abstract}

The Rice-Sneddon model for BaBiO$_3$ is a
nice model Hamiltonian for considering the properties
of polarons and bipolarons in a three-dimensional
oxide crystal.  We use exact diagonalization methods
on finite samples to study the stability and properties
of polarons and bipolarons.  Because polarons, when they
form, turn out to be very well-localized, we are able
to converge accurately our calculations for two-electron
bipolaron wavefunctions, accounting for the Coulomb
interaction without approximation.  Some of our results
are compared with and interpreted by reference to the
variational method of Landau and Pekar.  We calculate
both electronic and vibrational excitations of the
small polaron solutions, finding a single vibrational
state localized with the full symmetry of the polaron,
which has its energy significantly increased.
Both on-site (Hubbard) and long-range Coulomb
repulsion are included in the bipolaron calculation,
but due to the high degree of localization, the
long-range part has only a small influence.  For
a reasonable on-site repulsion $U$ equal to 
2 times the band width $W$, bipolaron formation
is significantly suppressed; there is a large window
of electron-phonon coupling where the polaron
is stable but the bipolaron decays into two polarons.

\end{abstract}
]
\narrowtext
\section{introduction}
The electron localized on an ion (or a few ions) 
can cause displacements of neighboring ions
from their positions in the crystal lattice. 
The quasiparticle formed by an electron 
and corresponding lattice displacements is called a {\em polaron}
\cite{Landau:1933}.  The polaron is a {\em small polaron} if
the electron is localized  on a few ions or a
{\em large polaron} otherwise \cite{Alexandrov:1995}. 
It is possible that the surrounding ions ``overscreen'' the negative
charge of the localized electron and attract another
electron to the same site. 
Two electrons bound in such way form a {\em bipolaron}
\cite{Alexandrov:1995,Mott:1990}. 
The important issue for bipolaron formation
is a quantitative analysis of the relative 
strength of the  electron-lattice interaction
responsible for two electrons being coupled
and the electron-electron (Coulomb) forces which    
try to break the bipolaron apart.  If the Coulomb repulsion 
between electrons 
exceeds some critical value the bipolaron is less stable than
two separated polarons.

In present paper we study the polaron and bipolaron formation in 
a model originally proposed by Rice and Sneddon
\cite{Rice:1981} for doped $\rm BaBiO_3$, such as
$\rm Ba_{1-x}K_xBiO_3$ (BKBO).   These materials are
nearly cubic perovskites, with a fairly simple set of
valence electron states, exhibiting superconducting
transition temperatures as high as 30K at optimal doping.
We choose this model
both because of its relation to physically important
materials (SrTiO$_3$ and WO$_3$ could be studied by a
closely related model with $T_{2g}$ $d$-states instead
of $s$-electron states)
and also because it makes a nice test model
for looking at the criteria for polaron formation and
the properties of polarons.   In this study, for
simplicity, we consider the hypothetical case of
an almost empty band, corresponding to KBiO$_3$, or BKBO with x=1.

Our principal results are (1) there is a sudden jump
at a critical coupling strength from a large (delocalized)
polaron to a very small, well-localized polaron; (2)
the bipolaron breaks apart
even at moderate strength of on-site Coulomb repulsion;
and (3) small polaron formation is accompanied by a 
characteristic formation of localized vibrations of 
shifted energy which may serve as a good experimental
signature.

\section{model Hamiltonian}

The Rice-Sneddon Hamiltonian is
\begin{equation}
H=H_t+H_{e-ph} + H_{ph} +H_C \>,
\label{eq:Hamiltonian}
\end{equation}
where $H_t$, $H_{e-ph}$, $H_{ph}$ and $H_C$ are , correspondingly,
hopping, electron-phonon, phonon, and Coulomb terms.
For the hopping term, only one bismuth $s$-orbital per cell
is used, with hopping only to nearest bismuth neighbors,
\begin{equation}
H_t  =  -t \sum_{\langle i,j \rangle \sigma}
\hat{c}^{\dagger}_{i\sigma} \hat{c}_{j\sigma}\>.
\label{eq:hopping}
\end{equation}
This has the usual cosine dispersion relation
with bandwidth, $W$, equal to $12t$.
We take the value of $t$ to be 200 $meV$
following band-structure calculations
\cite{Mattheiss:1988}.
We assume that the same model for electrons in BKBO can
be applied to all the region of K concentration,
because the antibonding Bi(6s)-O(2p) conduction band
in cubic BKBO is minimally affected by substitutional K doping
at the Ba sites\cite{Mattheiss:1988}.
First-principles LMTO calculations for different
phases of BKBO: $\rm KBiO_3$, $\rm K_{0.5}Ba_{0.5}BiO_3$
and $\rm BaBiO_3$ reveal a single band near Fermi energy
to be largely independent on potassium doping\cite{Mosley:1994}.

Hypothetical KBiO$_3$ in this model has no electrons.
Each Bi atom is in the 5+ ionization state with no
remaining valence electrons, whereas BaBiO$_3$ in this model
has a half filled band of Bi$^{4+}$ ions.  Real BaBiO$_3$
is insulating, and is most simply understood as having
alternating Bi$^{3+}$ and Bi$^{5+}$ ions.  The Bi$^{3+}$
ions have two valence electrons bound to them, and
repel the surrounding 6 oxygen atoms, which are attracted
to the more positive Bi$^{5+}$ ions.  A nice term for this
state is that it is a ``bipolaronic crystal.''  The
Rice-Sneddon Hamiltonian incorporates this possibility through
the terms
\begin{equation}
H_{e-ph}  =  \frac{g}{a} \sum_{i\sigma} \sum_{\alpha=1}^{3}
\Big[ u_{i-,\alpha}-u_{i+,\alpha} \Big] \hat{c}^{\dagger}_{i\sigma}
\hat{c}_{i\sigma} \>,
\label{eq:elph}
\end{equation}
\begin{equation}
H_{ph}  =  \sum_{i} \sum_{\alpha=1}^{3} \Big[
\frac{P^{2}_{i,\alpha}}{2M} +\frac{1}{2} M\omega_{0}^{2}
u_{i,\alpha}^{2} \Big]\>.
\label{eq:bareph}
\end{equation}
The notation $u_{i,\alpha}$ refers to the displacement of the
oxygen atom (labeled by \{ $i,\alpha$ \} ) which neighbors the
$i$-th Bi atom in the $\alpha$=x, y, z Cartesian direction.
Only displacements in the direction $\hat{\alpha}$ 
of the bond
are considered, because these are expected to dominate the
physics of polaron formation.  The notation $u_{i+,\alpha}$
means exactly the same as $u_{i,\alpha}$, whereas $u_{i-,\alpha}$
refers to the displacement of the oxygen atom which neighbors the
$i$-th Bi atom in the $\alpha= \rm -x, -y, -z$ Cartesian direction.
The atoms labeled by $(i-,\alpha)$ can also be labeled in
the form $(i',\alpha)$ by reference to the appropriate nearby
Bi atom $i'$.  Eq. (\ref{eq:elph}) contains the effect that
an ``inhaling'' of the negative oxygen ions around a central
Bi ion will raise the on-site energy of the
bismuth $s$-orbital, by amount equal
to $g$ per fractional displacement $u/a$ of each of the 6 surrounding
atoms.  However, this costs elastic energy  as given in
Eq. (\ref{eq:bareph}).  We take M as the oxygen atomic mass,
and $\omega_0$ to have the value
65 meV of a typical oxygen bond-stretching vibration.
(It corresponds to spring energy $M\omega_{0}^2a^2/2$
being $\approx$ 4 eV.)
Various values of $g$ will be used, in the physically expected
range of 1-3 eV.  The Bi-Bi interatomic
distance $a=4.28\AA$ will be used as the unit of length, and
the hopping parameter $t$ is used as the unit of energy.
Finally, there is a Coulomb interaction between electrons,
\begin{equation}
H_C   =  U\sum_{i} \hat{n}_{i\uparrow} \hat{n}_{i\downarrow} +
\sum_{i,j} \frac{e^2}{\varepsilon |r_i-r_j|}.
\label{eq:coulomb}
\end{equation}
Various values of $U$ comparable with $W=12t$ will be used
for the on-site (Hubbard) term in the Coulomb interaction,
and the value $\varepsilon = 5$ will characterize the long
range Coulomb repulsion.

The vacuum of this model corresponds to KBiO$_3$, with
no electrons, and only zero-point vibrational energy
$3N\hbar\omega_0/2$.  We will see that shifts of the zero
point energy are not very important, so we can ignore this
term and define the vacuum as having energy zero.
For the half-filled case, the ground state (for not too
strong Coulomb repulsion compared to electron-phonon
coupling) has a period-doubling distortion of the oxygens.
Alternate bismuth sites have the oxygens ``breathing'' either
in or out, and electron charge either diminished or increased
from the average of one electron per site.  This corresponds
to the experimental insulating state of BaBiO$_3$, and
gives a new vacuum into which carriers can be introduced by
doping.  Neglecting Coulomb repulsion, this regime was
was studied numerically by Yu {\sl et al.} \cite{Su:1990},
who found, in agreement with experiment, that the insulating
gap persisted for a wide range of K-doping. 
In d=2, this model was studied by Prelov\v sek {\sl et al.} 
using an adiabatic treatment of the
phonon degrees of freedom and the Hartree approximation
for the Hubbard term \cite{Prelovsek}.

\section{polaron}

Our only approximation (apart from finite size errors which
are well-controlled) is the Born-Oppenheimer (adiabatic)
treatment of the vibrations.  Inserting one electron into
the empty-band vacuum, and letting the oxygen atoms
have some fixed distortion pattern $\{u\}$, we look for the
lowest energy one-electron state with wavefunction
\begin{equation}
\psi_{\uparrow}(\{u\})=
\sum_i a_i(\{u\})c^{\dagger}_{i,\uparrow}|\rm{vac}\rangle\>,
\label{eq:1elpsi}
\end{equation}
where $a_i(\{u\})$ are the site amplitudes of the 
electron wave function. Later the 
dependence of $a_i$ {\sl etc.} on the parameters $\{u\}$ 
will be implicit and not explicitly designated.
This electron state has energy $\epsilon_0(\{u\})$,
measured relative to the bottom of the band,
$\epsilon_0(\{0\})$.  The total energy is this
plus the elastic energy $\langle H_{ph}(\{u\})\rangle$.
Then we vary the displacements $\{u\}$ looking for the
absolute minimum total energy.  If the coupling constant $g$
is small, the minimum occurs at $\{u\}=\{0\}$ and has
total energy 0.  This corresponds to a large polaron
solution, which in adiabatic approximation is just
an electron in the bottom of the band of the undeformed
crystal.  If we were to include the non-adiabatic
coupling of this electron with virtual phonons, there
would be an alteration of the mass and energy of this
electron.  Specifically, the energy shift would be
\begin{equation}
\Delta \epsilon_0 = \sum_{j,k} \frac{
|M_{0j}^k|^2}{\epsilon_0 -\epsilon_j -\hbar\omega_k}\\
\label{eq:selfen}
\end{equation}
in terms of the one electron energies $\epsilon_j$
and the phonon energies $\hbar\omega_k$ of the
unperturbed band.  This sum can be evaluated as follows,
\begin{eqnarray}
\frac{\Delta \epsilon_0 }{t}& = &\left( \frac{g}{t}\right)^2
\frac{\hbar \omega_0}{2M\omega_0^2 a^2}
\left[ 1- S\left(\frac{\hbar\omega_0}{4t}\right)\right],
\label{eq:selfen1}\\
S(x) & = & \frac{1}{N}
\sum_{\vec{Q}}\frac{x}
{x+f({\bf Q})} \>, 
\label{eq:selfensum}
\\
f({\bf Q}) & = & sin^2(\frac{Q_x a}{2})+sin^2(\frac{Q_y a}{2})
+sin^2(\frac{Q_z a}{2}) \>.
\label{eq:sinsum}
\end{eqnarray}
For our choices of $\omega_0$ and $t$, the ratio
$\hbar\omega_0 /4t$ is 0.081 and the sum $S$
in Eq. (\ref{eq:selfensum}) is 0.05.
Thus the self-energy shift of the large polaron
is $\approx 4 \times 10^{-3} \times (g/t)^2$
which turns out to be small compared to the energies
that we will find for the small polaron regime.
Thus we can safely ignore the non-adiabatic effects.
By a similar argument (which we will explain in more
detail later) the zero-point contribution to the
elastic energy can be ignored, and our elastic 
contribution to the small polaron energy is just
the second term of Eq. (\ref{eq:bareph}).

\begin{figure}
\centerline{\psfig{file=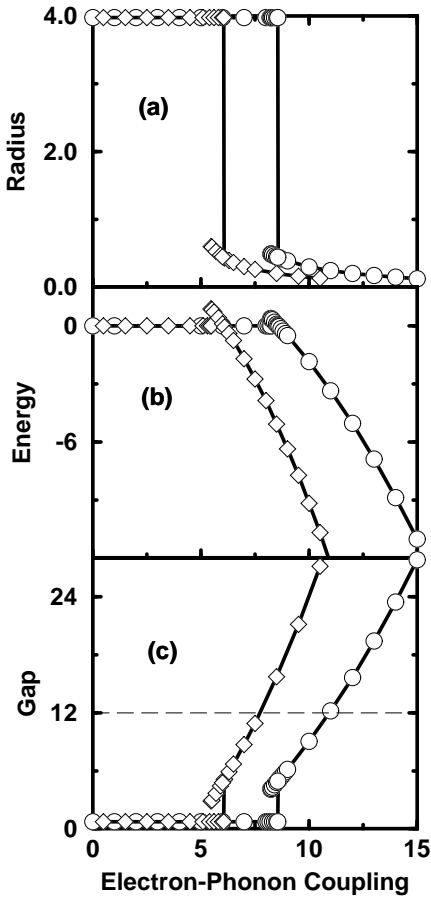,angle=-90,width=0.75\linewidth}}
\caption{Calculated properties of a polaron: (a)
radius $r_P /a$, (b) total energy per electron $E_{\rm tot}/t$,
and (c) the gap $(\epsilon_1-\epsilon_2)/t$ in the electron spectrum
as a function of electron-phonon coupling $g/t$.
The circles give results for the single-electron polaron
and the diamonds give the radius, half the total energy, and the
electronic gap for the bipolaron.
The Coulomb repulsion is
omitted for the bipolaron ($U=U'=0$).
}
\label{fig:nocoul}
\end{figure}

To evaluate the one-electron energy $\epsilon_0(\{u\})$
for the distorted lattice requires a finite size system,
which we choose to be an orthorhombic cell (our ``supercell'')
with $N=N_1 \times N_2 \times N_3$ Bi atoms on a cubic
lattice, and 3$N$ oxygens on the Bi-Bi bonds, and
periodic boundary conditions.  The Lanczos technique 
\cite{Cullum:1985} was used for finding the
ground state energy and a few lowest excited states
of the Hamiltonian (\ref{eq:Hamiltonian}), and
conjugate gradient minimization was used to find
the optimum values of the oxygen displacements $\{u\}$.
Beyond a critical value $(g/t)_c$=8.57 it becomes
favorable for oxygens to distort and form a localized
small polaron state.  We define the location 
$\vec{r}_0$ and radius $r_P$ of the polaron by
\begin{equation}
\vec{r}_0=\sum_i |a_i|^2 \vec{r}_i \>,
\label{eq:location}
\end{equation}
\begin{equation}
r_P^2 =\sum_i |a_i|^2 (\vec{r}_i-\vec{r}_0)^2 \>.
\label{eq:radius}
\end{equation}
The radius of the polaron at the transition
is 0.49 in units of Bi-Bi distance, that is, it
is very well-localized with 90\% of the electron density
concentrated on one site.  As $g$ increases beyond the
critical value, the radius further shrinks, and the binding
energy rapidly increases to values of order $t$ and bigger.

Our results are plotted in Fig. \ref{fig:nocoul}.
For $g/t$ less than the critical value, the radius
is shown as a finite number, $\approx 4a$, reflecting the
finite size of the cell; the actual radius is infinite.
For values of $g/t$ slightly less than critical,
our minimization proceedure locates a metastable
small polaron solution with a small positive energy,
which is shown in Fig. \ref{fig:nocoul} as
a small hysteretic region. 

\begin{figure}
\centerline{\psfig{file=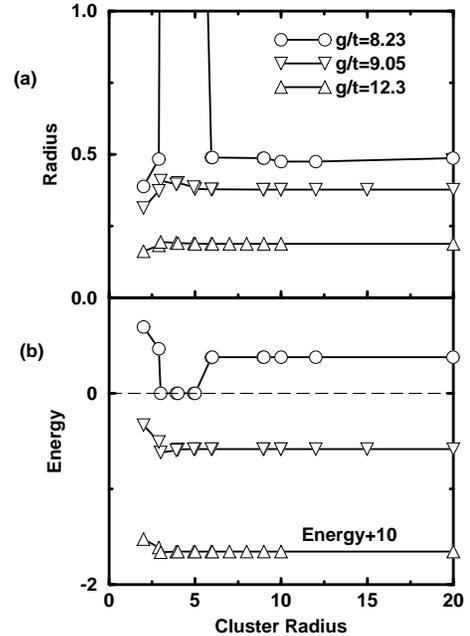,angle=-90,width=0.75\linewidth}}
\caption{Tests of finite-size errors for different values of electron-phonon
coupling constants. The cluster radius is in the units of the Bi-Bi
distance. The clusters used are $N_0 \times N_0 \times
N_0$ and $(N_0-1)\times N_0\times (N_0+1)$,
out to a maximum of $N_0$ of 20. The cluster radius is
$N_0$ in the former case or $N_{0}^{1/3}
(N_{0}^{2}-1)^{1/3}$ in last case.
}
\label{fig:sizerr}
\end{figure}

Because the small
polaron is so well-localized, the error in our
calculation due to the finite size supercell is
easy to control.  To test this, we have varied
the size of the supercell from a minimum of $2 \times
2 \times 2$ to maximum of $20 \times 21 \times 22$.
The results are shown in Fig. \ref{fig:sizerr}. 
The polaron radius and
total energy are insensitive to cluster size if the number 
of Bi-atoms is  $\agt 200$.
Near the  transition, for cluster size not too big,
the transition onset varies with cluster size. 
The total energy always diminishes with increase of cluster
size until it becomes independent of cluster size.
At $g/t$  far enough from the critical value
the results are almost the same
for all the clusters sizes.  The results of
Fig. \ref{fig:nocoul} have no noticeable size
dependence.

The non-adiabatic corrections to the large polaron energy
are 0.0064$t$ at the transition point, which gives
an unimportant correction to the critical value of $g/t$.
The small polaron solution is $2N$-fold degenerate: it
can form at any of the $N$ Bi sites, with either spin.
In this paper we ignore another non-adiabatic effect,
the weak vibration-assisted tunnelling which lifts the
translational degeneracy to make a narrow band with
only spin degeneracy remaining.

\begin{figure}
\centerline{\psfig{file=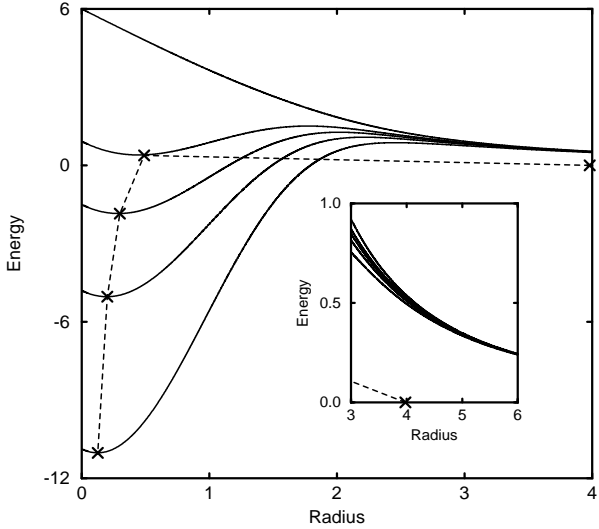,angle=-90,width=0.99\linewidth}}
\caption{Total energy {\sl versus} polaron radius calculated in
Landau-Pekar approximation for electron-phonon coupling
constants  $g/t$=0., 8.23, 10., 12., and 15.
from top to bottom. The energy minima for $g/t$= 10.,
12., and 15. correspond to energies of stable localized states.
The extremum
at $g/t$=8.23 corresponds to a metastable localized state.
Results from finite cluster diagonalization
are shown as X's.
For numerical comparison see Table 1.
The inset expands the large $r_P$ part of the same curves.
}
\label{fig:LP}
\end{figure}

We now compare our numerical results with analytic
results obtained by a variational method
introduced by Landau and Pekar (LP)
\cite{Pekar:1946,Miyake:1976}.
The electronic wave function is
chosen to have Gaussian form
\begin{equation}
a_i = C_0(\beta) \exp(-\beta^2(\vec{r}_i-\vec{r}_0)^2/2)\>,
\label{LP}
\end{equation}
where $C_0$ is the normalization constant and
$\beta$ is the variational parameter which we call the
LP parameter. 
There are now two sequential minimizations to
perform \cite{Mahan:1990}.  First for fixed $\beta$
the optimum displacements $\{u(\beta)\}$ are found.
Then these are used to evaluate the trial total
energy $E(\beta)$, and a second minimization is performed
to find the optimum $\beta$.  We find analytic formulas for
$E(\beta)$ and the polaron radius $r_P(\beta)$,
\begin{eqnarray}
E(\beta) &=& -6t q^{1/4} \frac{\theta_2 (q)}{\theta_3(q)} 
          -3 \frac{{\rm g}^2}{M\omega_0^2} \frac{\theta_3(q^2)^2}
{\theta_3(q)^6} 
\nonumber\\
& \times &
[ \theta_3(q^2) - q^{1/2} \theta_2 (q^2)] \>,\\
\label{eq:enLP}
r_P(\beta)^2 &=& 6 q \theta_{3}' (q)/\theta_3(q) \>,
\label{eq:radLP}
\end{eqnarray}
where $q={\rm exp}(-\beta^2)$ and 
$\theta_i(q)= \theta_i(0,q)$ are Jacobi's theta functions
\cite{Abramowitz}.
$\theta_{3}' (q)$ is the derivative of $\theta_{3} (q)$.
These equations define an implicit function $E(r_P)$ 
which is plotted in Fig. \ref{fig:LP} for the values 
$g/t$ equal to 0., 8.229, 10., 12. and 15. 
X-marks indicate the exact results from our finite cluster 
calculations.
The agreement is remarkable, especially, for the values of
minimal total energy (see Table 1). 
The LP solution gives a smaller value of the 
polaron radius. 
The variational solution for an infinite system agrees with
the exact solution on finite clusters in finding a first-order
large to small polaron transition, with no intermediate regime
of large polarons 
exist. Figure \ref{fig:LP} explains the hysteresis  
found in the numerical results obtained by exact-diagonalization.
For a small range of $g$ just below the critical value 
the small polaron state is locally stable but separated
by an energy barrier from the global (delocalized) minimum.
The numerical solution follows this
metastable branch until it disappears.    

\begin{table}
\caption{
Comparison of results obtained by Landau-Pekar
(LP) variational method and by exact diagonalization of finite clusters
(cluster). The electron-phonon coupling constants are in units of
bandwidth ($W=12t$). The first row in the table corresponds to the critical
coupling constant at which a metastable
localized state occurs. The total energies $E_{\rm total}^{\rm LP}$ and
$E_{\rm total}^{\rm cluster}$ are in units of $t$. The polaron radius is
given in units of the Bi-Bi distance.
Cluster calculations give a slightly smaller total energy and
larger radius. The phonon spectrum is perturbed by the
transition from delocalized state to localized (see text). In LP
approximation only one mode, $\omega_{\rm max}^{\rm LP}$ changes from its
initial value of $\omega_0$. In the cluster calculations, one mode,
$\omega_{\rm max}^{\rm cluster}$, is well-separated from the others.
}
\begin{tabular}{ccccccc}
 $g/W$  & $E_{\rm total}^{\rm LP}$ & $E_{\rm total}^{\rm cluster}$
   & $r_p^{\rm LP}$ & $r_p^{\rm cluster}$ & $\omega_{\rm max}^{\rm LP}/\omega_0$
   & $\omega_{\rm max}^{\rm cluster}/ \omega_0$\\ \tableline\tableline
0.686  &  0.398 &  0.377 & 0.448 & 0.490 & 1.176 & 1.163 \\ \tableline
0.833  & -1.848 & -1.853 & 0.290 & 0.299 & 1.075 & 1.065 \\ \tableline
1.000  & -5.040 & -5.041 & 0.198 & 0.200 & 1.035 & 1.028 \\ \tableline
1.250  & -11.03 & -11.03 & 0.126 & 0.126 & 1.014 & 1.011 \\
\end{tabular}
\end{table}

\section{Electronic and vibrational excitations of the polaron}

An advantage of the exact diagonalization method is that is
enables an equally good and easy calculation of electronic
excitations of the Franck-Condon type where the lattice
distortion is frozen in place.  We simply examine the next
higher-lying eigenstates without further alteration of
the parameters $\{u\}$.  In the range of parameter space we
have explored, we have not encountered a second bound state
in the polaronic well.  The electronic spectrum has a gap,
and the minimum energy electronic excitation is a delocalized
state.  The energy of this transition, denoted the ``gap''
energy, is plotted in Fig. \ref{fig:nocoul}.c.
At the onset of polaron formation, the gap has a
value $5.00t$ which increases rapidly for larger
coupling constants.  There is a small but noticeable
finite size error in the gap calculation since the
lowest electronic excited state is extended to infinity,
but cut off at the supercell boundary in our work.

When a small polaron is formed, the interaction
between the localized electron and the lattice
vibrations can cause both a renormalization of
the electron energy and of the phonon energy.
Referring to Eq. (\ref{eq:selfen}), it is clear that the
gap, or minimum value of $\epsilon_0 -\epsilon_j$
in the denominator, makes a change in the electron self-energy
shift relative to the one already calculated
for the delocalized large polaron, probably reducing the 
shift because of the larger denominator (although matrix
element changes need to be considered also.)  However,
since the shift is certainly small compared to the gap
itself (of order $t$), this effect can be neglected.

A more interesting effect is the change in the local
vibrations near the localized electron.  If we know
the one-electron energies $\epsilon_j$
and the corresponding states $|j\rangle$
at the optimal set of displacements $\{u_0\}$,
then standard perturbation theory for small displacements
around these displacements gives
\begin{eqnarray}
\Delta\left[\frac{\partial^2 E_{{\rm tot}}}
{\partial u_{\ell,\alpha} \partial u_{\ell',\beta}}\right]
& = & \sum_{j} \frac{
\langle 0| V_{\ell,\alpha}|j\rangle
\langle j| V_{\ell',\beta}|0\rangle
}{\epsilon_0 -\epsilon_j}\>, 
\label{eq:chspr}
\\
V_{\ell,\alpha} & = & 
\frac{\partial H_{e-ph}}{\partial u_{\ell,\alpha}} \>.
\nonumber
\end{eqnarray}
This equation omits terms containing second derivatives of
$H$ because they are consistently omitted in our model.

The LP approximation gives a particularly simple
solution to this problem.  Since we don't have a complete
set of states in this approach, instead, we find the 
energy as a general function of the displacements $\{u_0\}$
and the LP parameter $\beta$,
\begin{equation}
E(\vec{u},\beta)=\frac{1}{2} \vec{u}^{\dagger} \cdot \hat{A}
\cdot \vec{u} + \vec{L}^{\dagger}(\beta) \cdot \vec{u} +f(\beta)\>,
\label{eq:LPen}
\end{equation}
where $\vec{u}$ is the $3N$-vector displacement, $\hat{A}$
is the bare force constant matrix (which is a constant times
the unit matrix in our model), $\vec{L}$ is the force on the
atoms caused by the localized electron, and $f$ is the localized
electron hopping energy.  Expressions for $\vec{L}$ and $f$
are easy to derive.  Straightforward linear algebra leads
to expressions for the optimum values $\{u_0\}$ and $\beta_0$.
We then Taylor-expand Eq. (\ref{eq:LPen}) to second order for
small deviations $\{\delta u \}$ and $\delta \beta$ around
the optimum values.  Finally, for fixed deviations
$\{\delta u\}$ the optimum value of $\delta \beta$ is
found.  Inserting this into the Taylor expansion,
the modified force constant matrix $\hat{A}'$ is found,
\begin{equation}
\hat{A}'=\hat{A}+\frac{\vec{L'}\vec{L'}^{\dagger}}
{\vec{L}^{\dagger}\cdot\hat{A}^{-1} \cdot\vec{L''} - f''} \>.
\label{eq:LPspr}
\end{equation}
The primes on the right hand side of Eq. (\ref{eq:LPspr})
denote derivatives by $\beta$.  Note that the alteration
of the force constant matrix in LP approximation is
factorizable, and since $\hat{A}$ is proportional
to the unit matrix, only one eigenvalue is altered, 
the corresponding eigenvector being proportional
to $\vec{L'}$.  The static displacements $\{u_0\}$
in LP approximation are given by $-\hat{A}^{-1}\vec{L}(\beta_0)$.
Thus in LP approximation, one vibrational eigenvector
splits off from the degenerate frequency $\omega_0$,
shifting to higher energy, and having an eigenvector
proportional to the derivative of the static
displacements by the LP parameter $\beta$.
The symmetry of this mode is identical to the
symmetry of the static displacement.

\begin{figure}
\centerline{\psfig{file=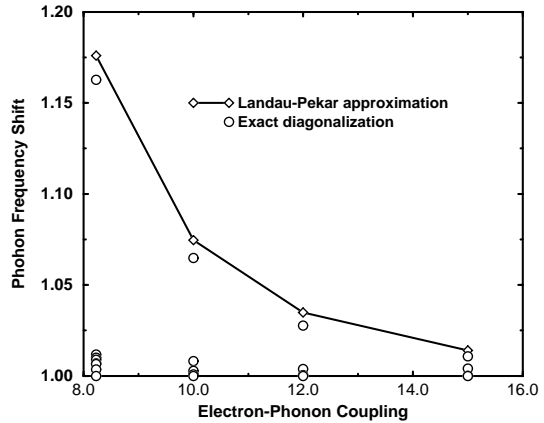,angle=-90,width=0.9\linewidth}}
\caption{Phonon frequency shifts, $\omega_{ph}^{i}/\omega_0$, at
coupling constants $g/t$=8.23, 10., 12., and 15. for cluster
size $5\times 6\times 7$. The phonon modes (circles)
are distributed from $\omega_0$ to $\omega_{\rm max}$, the last
being well separated from others. Results from the Landau-Pekar
approximation are shown by diamonds.
}
\label{fig:vib}
\end{figure}

We have also made an exact calculation of the modified
vibrational spectrum using finite clusters, and the
answers, shown in Fig.\ \ref{fig:vib}, 
and also in Table 1, agree nicely
with the LP approximation, but in addition to the
one strongly altered frequency, a few other frequencies
are pulled weakly above the unperturbed frequency
$\omega_0$.  

Thus we expect that a characteristic signature of the
small polaron state should be a localized vibrational
mode whose symmetry copies that of the polaronic
distortion, that is, the symmetry is the same as
the point symmetry in the crystal of the ion on
which the polaron is centered (full cubic symmetry
$A_1$ in our case.)  Such modes might be measureable
by Raman scattering using a laser which is resonant
with an electronic transition of the polaron.
Also they might appear as side-bands on the
electronic polaron absorption spectrum.

\section{bipolaron}

\begin{figure}
\centerline{\psfig{file=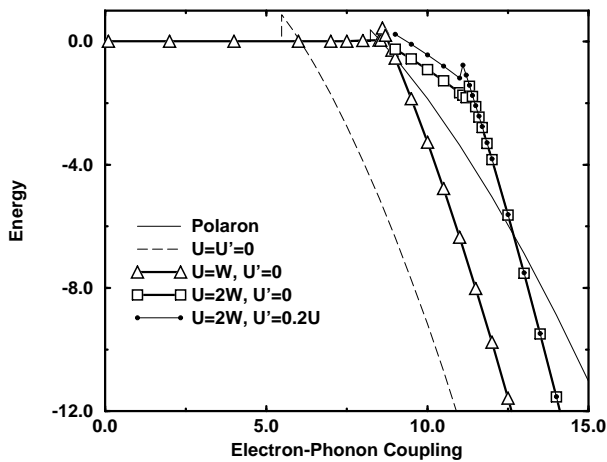,angle=-90,width=0.99\linewidth}}
\caption{Total energy (per electron) of the bipolaron for different
values of Coulomb repulsion.
The supercell size is $5\times 6\times 7$.
At $U=2W$ for $g/t$ between 8.6 and 11.3, the
total energy corresponds to two small polarons separated
as far as possible in the cell;  because the cell is not
infinitely large, there is Coulomb repulsion which raises
the energy above the isolated polaron energy shown as
the thin curve (identical to Fig. \protect\ref{fig:nocoul}).
}
\label{fig:2en}
\end{figure}

We now ask what happens in our model when a
second electron is added.  If we neglect the
Coulomb interaction, the answer is that two
spatially separated polarons are unstable relative
to formation of a singlet bipolaron state in which
both electrons are on the same site.  If we allow 
no further lattice relaxation beyond the single
electron polaron, then the energy of the bipolaron
is already lower than two separated polarons 
because the (negative) electronic eigenvalue is
doubled but the positive lattice strain energy is
unchanged; additional
lattice relaxation will occur only if it lowers the
energy, and since there are now two electrons
exerting each force $g$ on the neighboring oxygens,
there will be additional relaxation.  Results
are shown in Fig.\ \ref{fig:nocoul} where we
plot the total energy per electron. The
critical coupling for bipolaron formation
is $g/t$=6.08, significantly less than for
polaron formation which starts at $g/t$=8.57.
At onset of bipolaron formation, the radius and the electronic gap
(0.44$a$ and
5.00$t$, respectively)
are 
approximately the same as
for polaron formation at its onset,
but at equal
values of $g$ the bipolaron is smaller and has
a larger gap.

Of course it is unrealistic to ignore the Coulomb
repulsion which will act in the direction of
destabilizing the bipolaron.  Our model permits us
to make exact (finite size) calculations for the
bipolaron by solving the appropriate two-particle
equation, that is, finding the exact two-particle
wavefunction
\begin{equation}
\Psi(\{u\})=
\sum_{i,j} a_{i,j}(\{u\})c^{\dagger}_{i,\uparrow}
c^{\dagger}_{j,\downarrow}|\rm{vac}\rangle.
\label{eq:2elpsi}
\end{equation}
This calculation is  of course far more demanding than
the corresponding polaron case, Eq. (\ref{eq:1elpsi}),
because it requires on each step of the minimization procedure
finding the smallest eigenvalues
of a $N^2 \times N^2$ matrix rather than a $N \times N$ 
matrix.  Our Lanzcos algorithm has allowed us
to calculate for $N$ as large as $5 \times 6 \times 7$.
Because the bipolaron turns out to be again well-
localized, the finite system size does not cause
a noticeable error.  Also for the same reason,
the long-range Coulomb repulsion is not
very important.

\begin{figure}
\centerline{\psfig{file=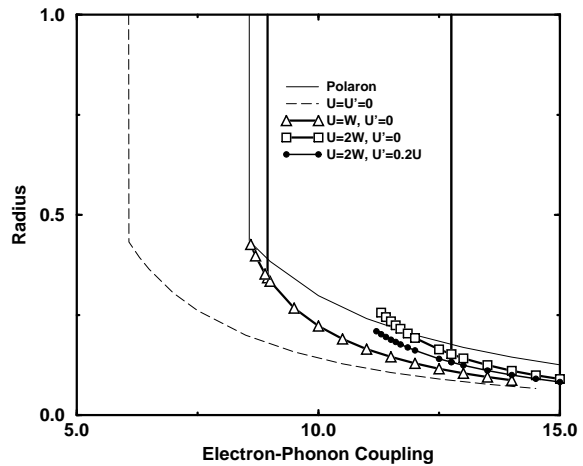,angle=-90,width=0.99\linewidth}}
\caption{Bipolaron radius as a function of $g/t$
for different values of Coulomb repulsion.
Vertical lines indicate stability limits where bipolarons
(with $U \protect{\alt} W$) decay into isolated small polarons, or
into isolated large polarons when $U \protect{\agt} W$.
At $g/t$ less than the stability point, radii of
metastable bipolaron states are shown.  The domain
of metastability is artificially enhanced by finite
size effects.
}
\label{fig:2rad}
\end{figure}

\begin{figure}
\centerline{\psfig{file=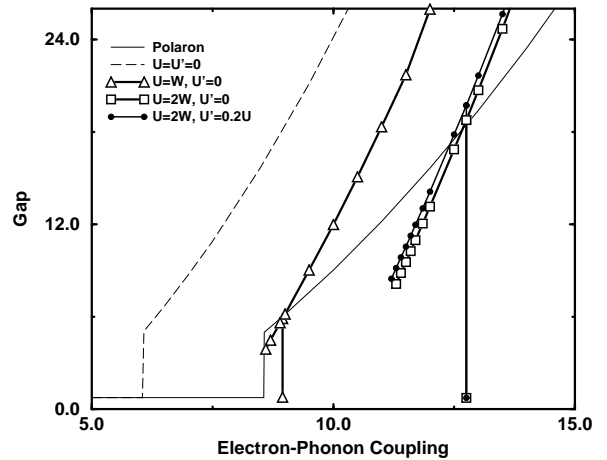,angle=-90,width=0.99\linewidth}}
\caption{The gap in the electron spectrum as a function of $g/t$.
Due to finite size of cluster the gap is finite even in
delocalized state.
}
\label{fig:2gap}
\end{figure}

Results for the bipolaron radius, energy, and
excitation gap are shown in Figs.\
\ref{fig:2en}, \ref{fig:2rad}, and \ref{fig:2gap}.
The onsite Coulomb strength $U$ (see Eq. (\ref{eq:coulomb}))
is estimated as $e^2/2r_{\rm Bi}$, while the long-range
Coulomb interaction is characterized by a parameter
we call $U'=e^2/\varepsilon a$.  For $r_{\rm Bi} = 0.5$ 
and $\varepsilon =5$ one finds $U$ = 3.5 eV = 1.46$W$
and $U'$ = 0.7 eV = 0.29$W$, where the bandwidth 
$W$ is $12t$ or $\approx 2.4$ eV.  Our calculations
are shown for the cases $U/W$= 1 and 2 with $U'$=0,
and for $U/W$= 2 with $U'$= 0.2.
Fig.\ \ref{fig:2en} shows that on-site Coulomb repulsion
with the small value of $U=W$ destabilizes the bipolaron
until $g/t$ reaches 8.95, slightly beyond the stability
point $g/t=8.57$ for the polaron.  
A more physical choice of $U=2W$ prevents stable bipolaron
formation until $g/t \approx$ 12.8.
Our numerical solutions show bipolarons winning over separated
polarons at somewhat smaller values of coupling constants
($g/t=8.60$ for $U=W$ and $g/t=11.3$ for $U=2W$.)  This is
a finite size error caused by the fact that in our cell,
two separated polarons are only separated by half the
cell diagonal, and have a small repulsion which artificially
raises their energy, favoring bipolarons for smaller coupling
than would be the case in a very large cell.  But since
we have already an accurate answer for the energy of two
well-separated polarons, our numerics tells us the true
stability point.  As is seen in Fig.\ \ref{fig:2en}, there
is also a small hysteretic regime indicating that the 
transition from separated polarons to bipolarons is first order.
As is seen in Fig.\ \ref{fig:2en}, adding the long-range
Coulomb term ($U'=0.2U$) causes a small additional postponement of
bipolaron formation, but hardly alters the properties
for values of $g/t$ where it is stable.

The radius shown in Fig.\ \ref{fig:2rad} is defined
exactly as in Eq. (\ref{eq:radius}) after defining
the probability $|a_i '|^2$ for finding an electron
on site $i$ in the usual way,
\begin{equation}
|a_i '|^2 = \sum_j |a_{i,j}|^2 \>.
\label{eq:2rad}
\end{equation}
Fig.\ \ref{fig:2rad} shows that
the radius of the bipolaron is increased by
increasing the value of $U$, but decreased  by
increasing the value of $U'$.  In  all cases 
the bipolaron has a smaller radius than the
polaron at the same value of $g$. 

An interesting feature shown in Fig.\ \ref{fig:2gap}
is that the gap is larger at the onset of bipolaron formation
in the presence of the long-range Coulomb repulsion,
presumably due to stronger localization of electrons. 
At fixed $g/t$ the Coulomb forces reduce the gap with
increasing $U$. 

%xxxxxxxxxxxxxxxxx
\section{summary}

Bipolaron formation
is strongly affected by Coulomb forces in a cubic
perovskite lattice. Due to Coulomb repulsion between two
electrons localized on the same site the onset of bipolaron
formation  can be postponed
and  polaron states are  energetically
favorable. 
The polarons and bipolarons formed in this lattice are
small and exist only above some critical value of
electron-phonon coupling. The transition from delocalized to
localized polaron state is discontinuous, with no intermediate-
size solution. 
This jump is not caused by  
finite-size errors and is present also in variational calculations
using Landau-Pekar approximation.  The total energy has hysteretic
behavior with metastable states occuring near the critical
coupling constant.  These metastable states could in principal
be observed, for example, by tuning the coupling constant with
applied pressure.
A gap opens in the electron spectrum
at the transition from delocalized
to localized polaron states, and new localized vibrational states
occur with energies increased above those of the undoped host.

\acknowledgements 
We thank V. Emery for discussions.
This work is supported by NSF Grant No. DMR 9417755.

\end{document}